\documentclass[lettersize,journal]{IEEEtran}
\usepackage{amsmath,amsfonts}
\usepackage{algorithmic}
\usepackage{array}
\usepackage[caption=false,font=normalsize,labelfont=sf,textfont=sf]{subfig}
\usepackage{textcomp}
\usepackage{stfloats}
\usepackage{url}
\usepackage{verbatim}
\usepackage{graphicx}
\def\BibTeX{{\rm B\kern-.05em{\sc i\kern-.025em b}\kern-.08em
    T\kern-.1667em\lower.7ex\hbox{E}\kern-.125emX}}
\usepackage{balance}
\begin{document}

\title{Coordinate-conditioned Deconvolution for Scalable Spatially Varying High-Throughput Imaging}
\author{Qianwan~Yang,~\IEEEmembership{Student Member,~IEEE},
Zhixiong~Chen,~\IEEEmembership{Student Member,~IEEE},
Jiaqi~Zhang,
Ruipeng~Guo,~\IEEEmembership{Student Member,~IEEE},
Guorong~Hu,~\IEEEmembership{Student Member,~IEEE},
and Lei~Tian,~\IEEEmembership{Member,~IEEE}%
\thanks{Manuscript created in December 2025. This project has been made possible in part by National Institutes of Health (R01NS126596) and a grant from 5022 - Chan Zuckerberg Initiative DAF, an advised fund of Silicon Valley Community Foundation. The authors thank the Boston University Shared Computing Cluster for providing computational resources. (Corresponding author: Lei Tian.)}%
\thanks{Qianwan Yang is with the Department of Electrical and Computer Engineering, Boston University, Boston, MA 02215 USA (e-mail: yaw@bu.edu).}%
\thanks{Zhixiong Chen is with the Department of Electrical and Computer Engineering, Boston University, Boston, MA 02215 USA (e-mail: zhixiong@bu.edu).}%
\thanks{Jiaqi Zhang is with the Department of Electrical and Computer Engineering, Boston University, Boston, MA 02215 USA (e-mail: jzhang@bu.edu).}%
\thanks{Ruipeng Guo is with the Department of Electrical and Computer Engineering, Boston University, Boston, MA 02215 USA (e-mail: rguo@bu.edu).}%
\thanks{Guorong Hu is with the Department of Electrical and Computer Engineering, Boston University, Boston, MA 02215 USA (e-mail: grhu@bu.edu).}%
\thanks{Lei Tian is with the Department of Electrical and Computer Engineering and the Department of Biomedical Engineering, Boston University, Boston, MA 02215 USA (e-mail: leitian@bu.edu).}%
}
\maketitle

\begin{abstract}
Wide-field fluorescence microscopy with compact optics often suffers from spatially varying blur due to field-dependent aberrations, vignetting, and sensor truncation, while finite sensor sampling imposes an inherent trade-off between field of view (FOV) and resolution. Computational Miniaturized Mesoscope (CM$^2$) alleviate the sampling limit by multiplexing multiple sub-views onto a single sensor, but introduce view crosstalk and a highly ill-conditioned inverse problem compounded by spatially variant point spread functions (PSFs). Prior learning-based spatially varying (SV) reconstruction methods typically rely on global SV operators with fixed input sizes, resulting in memory and training costs that scale poorly with image dimensions.
We propose \textbf{SV-CoDe} (Spatially Varying Coordinate-conditioned Deconvolution), a scalable deep learning framework that achieves uniform, high-resolution reconstruction across a 6.5 mm FOV. Unlike conventional methods, SV-CoDe employs coordinate-conditioned convolutions to locally adapt reconstruction kernels; this enables patch-based training that decouples parameter count from FOV size. SV-CoDe achieves the best image quality in both simulated and experimental measurements while requiring 10$\times$ less model size and 10$\times$ less training data than prior baselines. Trained purely on physics-based simulations, the network robustly generalizes to bead phantoms, weakly scattering brain slices, and freely moving \textit{C. elegans}. SV-CoDe offers a scalable, physics-aware solution for correcting SV blur in compact optical systems and is readily extendable to a broad range of biomedical imaging applications.
\end{abstract}


\section{Introduction}
\IEEEPARstart{F}{luorescence} microscopy is a fundamental tool for studying biological structures and dynamics. To move beyond benchtop instruments, miniaturized fluorescence microscopes (miniscopes) have emerged as a promising direction for in-vivo and wearable imaging~\cite{aharoni2019all}, but expanding their field of view (FOV) while maintaining micron-scale spatial resolution remains challenging~\cite{park2021review}.

\begin{figure}[t]
\centering\includegraphics[width=\linewidth]{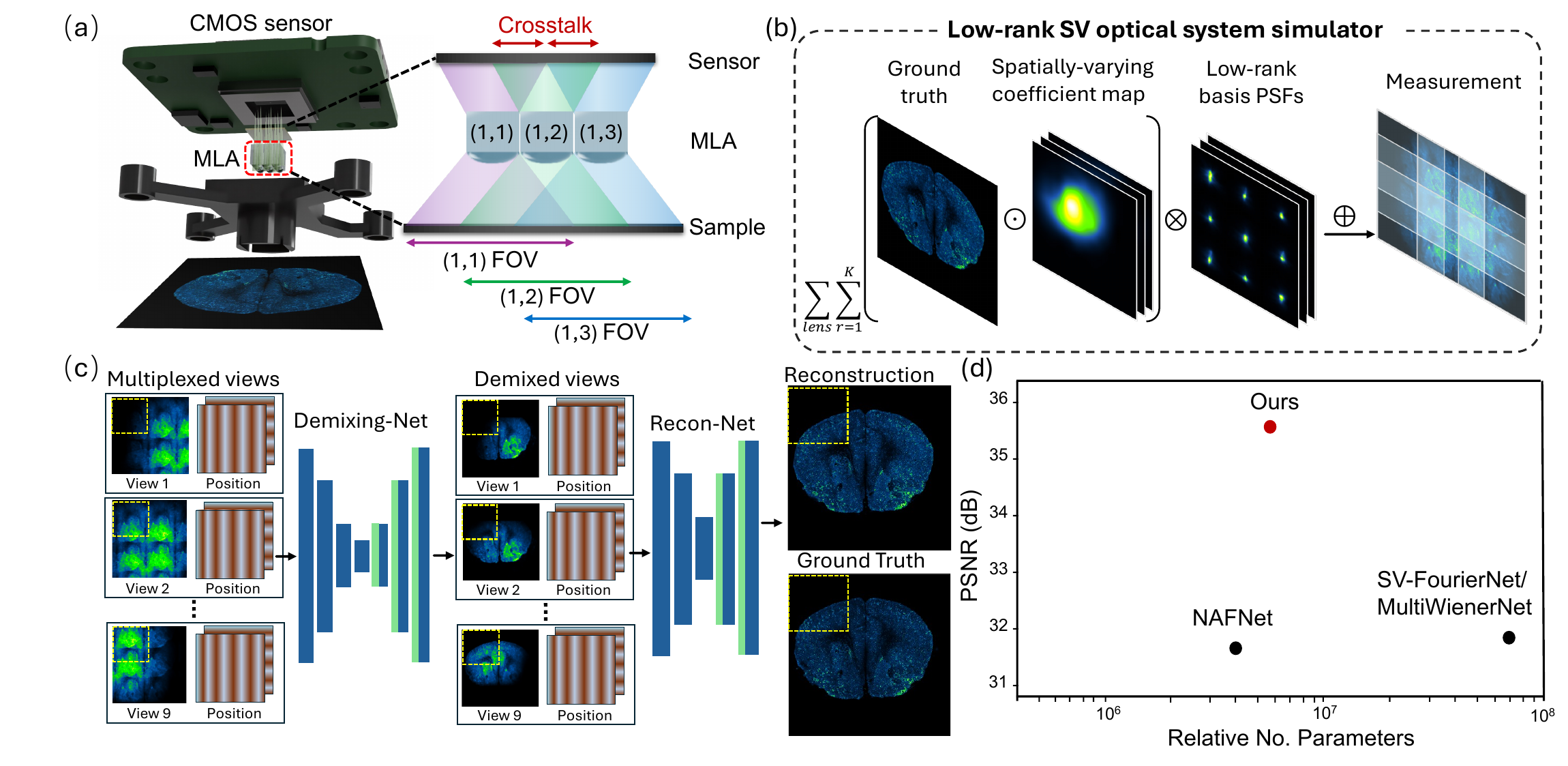}
\caption[System overview.]{\textbf{System overview.} (a) A multi-aperture miniscope uses an MLA as the sole imaging element; each microlens captures a distinct sub-FOV, and the sensor records an extended FOV with inter-view crosstalk. (b) Low-rank, SV forward model used for physics-based data synthesis. The simulated measurement is partitioned into multiplexed sub-views that serve as network inputs. (c) SV-CoDe pipeline: multiplexed sub-views and normalized spatial coordinates are cropped into patches and processed by a view-dependent Demixing-Net with CoDe blocks, followed by a Recon-Net that performs SV deconvolution to fuse views into a wide-FOV image. (d) SV-CoDe augments NAFNet with coordinate-conditioned spatial attention via a lightweight MLP, enabling SV reconstruction with minimal memory overhead and achieving higher PSNR with substantially fewer parameters than baseline methods.}
\label{fig:overview_svcode}
\end{figure}

Existing miniscope designs broadly fall into two categories. Lensless imaging enables extreme miniaturization by playing a phase mask close to the sensor, but the lack of focusing power leads to reduced contrast, and limited robustness to fluorescence background and scattering~\cite{boominathan2021recent,wu2024mesoscopic}. In contrast, multi-view or multi-aperture configurations retain focusing capability by capturing multiple sub-views, but achieving a large FOV often requires specialized optics~\cite{hu2024metalens} or sensor arrays~\cite{tanida2001thin}, which are costly and compromise compactness. Our lab developed the Computational Miniature Mesoscope (CM$^2$)~\cite{xue2020single,xue2022deep}, which follows the multi-view principle while preserving a compact form factor by placing a microlens array (MLA) directly in front of a single image sensor and intentionally allowing sub-views to overlap to maximize usable FOV (Fig.~\ref{fig:overview_svcode}(a)).

This optical simplicity shifts the burden from hardware to computation, rendering image reconstruction the primary performance bottleneck. The resulting inverse problem is highly ill-conditioned: strong view multiplexing and the periodic lattice of the MLA introduce substantial ambiguities in the measured data. Adding to this challenge, the PSF is inherently \emph{spatially variant} across the FOV due to field-dependent aberrations, vignetting, and truncation imposed by the finite sensor size. Consequently, conventional shift-invariant (SI) convolutional models fail to accurately invert the forward imaging process, often producing artifacts and pronounced resolution degradation toward the FOV periphery. Classical model-based approaches partially address these issues through iterative optimization, but their high computational cost and sensitivity to manually tuned regularization parameters limit scalability and robustness.

Recent deep learning approaches have therefore emerged as a promising alternative, offering improved reconstruction quality and speed for spatially-varying (SV) deconvolution. Learning-based SV deconvolution methods typically follow two paradigms. The first approximates spatial variance as a linear combination of SI operators~\cite{yang2024wide,yanny2022deep,wu2023real,tian2025deepinminiscope}. Although effective, these approaches typically require a global receptive field to synthesize SV filters across the sensor. As a result, model complexity scales with both the FOV and the degree of spatial variation (e.g., the effective rank of the SV operator), leading to high memory consumption and fixed input-size constraints that limit scalability for high-throughput imaging. 
The second paradigm makes the convolution itself coordinate-dependent~\cite{howard2024coordgate,liu2018intriguing,wei2024thin,fu2023field}, directly realizing SV filtering through lightweight coordinate conditioning. This approach enables local adaptation of reconstruction kernels without requiring a global receptive field and is therefore well suited for patch-based processing. However, it has not yet been fully explored in multi-aperture reconstruction, where SV arises jointly from sensor-wide field dependence and lens-dependent effects under view multiplexing. Addressing both forms of SV while preserving efficient, scalable patch-wise inference remains an open challenge.

Here, we introduce SV-CoDe, a lightweight and scalable framework that jointly performs view-aware SV demixing and deconvolution to address multiplexing-induced crosstalk and field-dependent blur, as illustrated in Fig.~\ref{fig:overview_svcode}(c).
SV-CoDe adopts a two-stage architecture consisting of view demixing followed by reconstruction, with both stages built from a shared CoDe block backbone based on an efficient NAF-style design~\cite{chen2022simple}. Each CoDe block replaces spatially invariant convolution with a coordinate-conditioned operation by introducing a lightweight MLP that maps coordinate-derived positional encodings to SV masks.

After each convolution, the predicted masks are applied multiplicatively to the feature maps, enabling position-dependent modulation that mirrors the coefficient maps in the SV forward model~\cite{yang2024wide} while introducing only negligible additional parameters. During inference, these masks depend solely on spatial coordinates and are therefore precomputed and cached, keeping runtime close to that of standard convolution. This design supports patch-based training while preserving global SV behavior, enabling large-FOV reconstruction with minimal memory overhead.

Trained solely on physics-based simulations using our previously developed low-rank SV forward model~\cite{yang2024wide} (Fig.~\ref{fig:overview_svcode}(b)) and validated experimentally, SV-CoDe achieves uniform, high-quality reconstruction across a 6.5-mm FOV while reducing memory usage by approximately $10\times$ compared to prior SV networks (Fig.~\ref{fig:overview_svcode}(d)).
Simulations show that the coordinate-gated backbone with positional encoding restores high spatial frequencies across the extended FOV and suppresses artifacts characteristic of SI models. Complementary experiments further demonstrate robust generalization across distinct regimes, including consistent resolution across the FOV, tolerance to wide variations in emitter density in particle phantoms, and transfer to weakly scattering brain slices and freely moving \emph{C.~elegans}. Together, these examples highlight SV-CoDe’s ability to capture the underlying SV structure of the imaging problem and generalize beyond the specific conditions encountered during training.

With minimal overhead and strong scalability, SV-CoDe provides a practical path to high-throughput, wide-FOV fluorescence imaging and is readily adaptable to other SV reconstruction problems.

\section{Method}
\subsection{Physics-based data simulation}
\noindent To bypass the need for extensive experimental data collection, we synthesize training data using a low-rank SV imaging model~\cite{yang2024wide}. Each simulated measurement is zero-padded to $4200{\times}4200$ pixels to compensate for sensor cropping, then partitioned into sub-views according to the chief-ray coordinates of the lenses in the MLA, yielding sub-views of size $2400{\times}2400$ pixels. During training, the multiplexed sub-views are cropped into overlapping patches of size $480{\times}480$ with a stride of $240{\times}240$ and stacked along the channel dimension to form a $480{\times}480{\times}9$ input to the demixing network, where the third dimension corresponds to the nine views. This procedure produces $16{\times}16 = 256$ patches per full measurement.

The training dataset combines naturalistic biomedical images (cells, vasculature, and brain sections) with synthetic fluorescent particle phantoms spanning wide ranges of emitter density, size, and intensity. To enhance robustness, noise is injected on the fly using a calibrated mixed Poisson–Gaussian model~\cite{yang2024wide}. Ground-truth images undergo background suppression and contrast normalization, followed by manual screening to ensure a high-quality fluorescence dataset with diverse sparsity levels and structural content. The base dataset contains approximately $1{,}307$ images, yielding $105{,}867$ patches in total. During training, one third of the patches are randomly sampled per epoch, corresponding to approximately $35{,}000$ patches.

Compared to our previously developed SV-FourierNet~\cite{yang2024wide} and MultiWienerNet~\cite{yanny2022deep}, the proposed patch-wise training strategy requires approximately $10{\times}$ fewer training pairs while achieving higher SSIM and PSNR across diverse test samples. In prior methods, training memory scales linearly with the measurement size, rendering end-to-end optimization prohibitive for large FOV. In contrast, our patch-based regimen decouples memory consumption from the FOV size, enabling scalable training on arbitrarily large FOV and making the approach readily extensible to other high-throughput imaging applications.

\subsection{SV-CoDe Network design}

\begin{figure}[t!]
\centering\includegraphics[width=\linewidth]{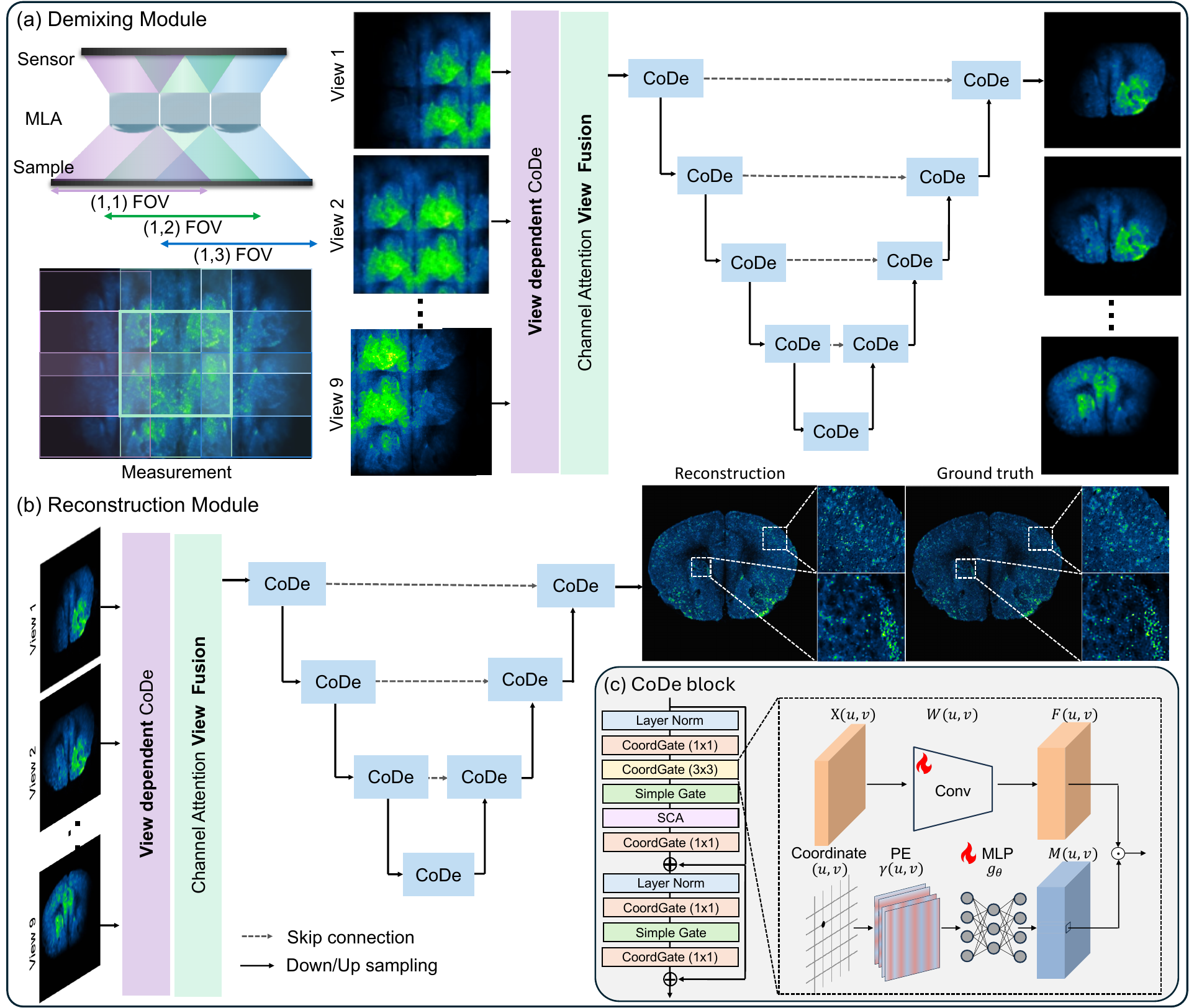}
\caption[Network design.]{\textbf{Network design.}
(a) \emph{Demixing module:} The raw measurement is partitioned into microlens sub-views using chief-ray coordinates. Each sub-view is processed by a view-dependent CoDe block, and the resulting features are passed to a U-Net composed of shared CoDe blocks.
(b) \emph{Reconstruction module:} The demixed sub-views are fed into the reconstruction network, which begins with a CoDe block to perform lens-dependent SV deconvolution. Channel attention then fuses information across views, followed by a second U-Net stack to produce a wide-FOV, high-resolution reconstruction.
(c) \emph{CoDe block:} The CoDe block replaces SI convolution with a coordinate-conditioned operation. As shown in the zoom-in, pixel coordinates \((i,j)\) are positional encoded and mapped by a lightweight MLP to a per-channel SV mask that modulates the post-convolution features.}
\label{fig:network_design}
\end{figure}

SV-CoDe is designed based on the physical SV forward model, in which spatial variance is expressed as a linear combination of SV convolutions parameterized by \emph{coefficient maps}. We approximate the inverse of this operator using a lightweight, coordinate-conditioned function, enabling view demixing and position-aware deconvolution with \emph{minimal} overhead.

The framework consists of two modules, \emph{Demixing-Net} and \emph{Recon-Net}, following a similar two-stage structure to~\cite{xue2022deep}. Raw measurements are partitioned into nine sub-views based on chief-ray coordinates and stacked, enabling local, patch-wise processing without requiring a global receptive field to aggregate sensor-wide multi-view information. \emph{Demixing-Net} operates on these multiplexed sub-view patches to suppress view crosstalk while preserving lens-specific content (Fig.~\ref{fig:network_design}\,(a)). The demixed sub-views are then passed to \emph{Recon-Net}, which performs SV deconvolution to correct lens-dependent aberrations and fuses information across views to produce a uniform, wide-FOV reconstruction (Fig.~\ref{fig:network_design}\,(b)).

Both modules follow the same design: (i) a \emph{view-dependent CoDe block} for lens-specific adaptation; (ii) channel attention with view-wise pooling to aggregate information across views; and (iii) a \emph{shared} UNet encoder–decoder with additional CoDe blocks for SV view fusion. This coupling of lens-specific demultiplexing and aberration correction and shared SV fusion reduces memory and stabilizes optimization.

The core building blocks for SV-CoDe is the CoDe block. It follows the NAFBlock topology~\cite{chen2022simple} but \emph{replaces every convolution in the NAFBlock with the coordinate gate (CoordGate)~\cite{howard2024coordgate}}. This preserves the efficiency of NAFBlock while enabling SV processing via a coordinate-conditioned mask, as shown in Fig.~\ref{fig:network_design}(c).

Given a feature map $F\in\mathbb{R}^{C\times H\times W}$ and normalized pixel coordinates $\mathbf{r}=(u,v)\in[-1,1]^2$, we form a positional encoding~\cite{mildenhall2021nerf} with $L$ frequency bands per axis:
\begin{equation}
\begin{aligned}
\gamma(\mathbf r)=\bigl[\dots,&\sin(2^k\pi u),\cos(2^k\pi u),\\
                        &\sin(2^{k+1}\pi v),\cos(2^{k+1}\pi v),\dots\bigr]_{k=0}^{L-1}.
\end{aligned}
\end{equation}

A lightweight MLP \(g_{\theta,\ell}\) generates a per-channel, coordinate-conditioned mask using \emph{only} the positional encoding. For each microlens \(\ell\), the mask is defined as
\[
M_\ell(\mathbf r)=g_{\theta,\ell}\bigl(\gamma(\mathbf r)\bigr).
\]
In the first \emph{view-dependent CoDe block}, the MLP is lens-specific to capture per-lens characteristics. In all subsequent CoDe blocks within the U-Net, a single shared MLP \(g_\theta\) is used across lenses to reduce memory consumption while retaining spatially varying modulation.

The mask is applied after convolution to modulate the feature maps, thereby realizing SV deconvolution:
\begin{equation}
\operatorname{CoordGate}(F; M_\ell)= M_\ell \odot (W * F),
\label{eq:coordgate}
\end{equation}
where $W$ denotes the convolution kernel, $*$ the convolution operator, and $\odot$ the Hadamard product. This formulation retains the structure of standard convolution while enabling position-dependent modulation through the coordinate-conditioned mask.

Because $M_\ell(\mathbf r)$ depends only on spatial coordinates, the masks can be precomputed after training, and cached at inference. Each CoDe layer therefore reduces to a standard convolution followed by element-wise gating, incurring negligible additional latency relative to a conventional NAFBlock.

Both stages are supervised jointly using a composite loss,
\begin{equation}
\mathcal{L}=\mathcal{L}_{\text{demix}}+\mathcal{L}_{\text{recon}},\qquad
\mathcal{L}_{(\cdot)}=\alpha\!\left(1-\mathrm{SSIM}\right)+\beta\,\mathrm{MSE},
\end{equation}
where SSIM promotes structural fidelity and MSE enforces intensity accuracy. The weights $\alpha,\beta\!\ge\!0$ balance the two terms and are set to $1.0$ in all experiments. Gradients are propagated end-to-end through both Demixing-Net and Recon-Net.

SV-CoDe is implemented in PyTorch and trained on simulated data using Adam with a cosine-annealing learning-rate schedule (batch size 4). Training a full model requires approximately 48 hours on a single NVIDIA A40 GPU.

\section{Results}
\subsection{SV-CoDe Achieves State-of-the-Art Reconstruction Quality}
\begin{figure}[t!]
\centering\includegraphics[width=\linewidth]{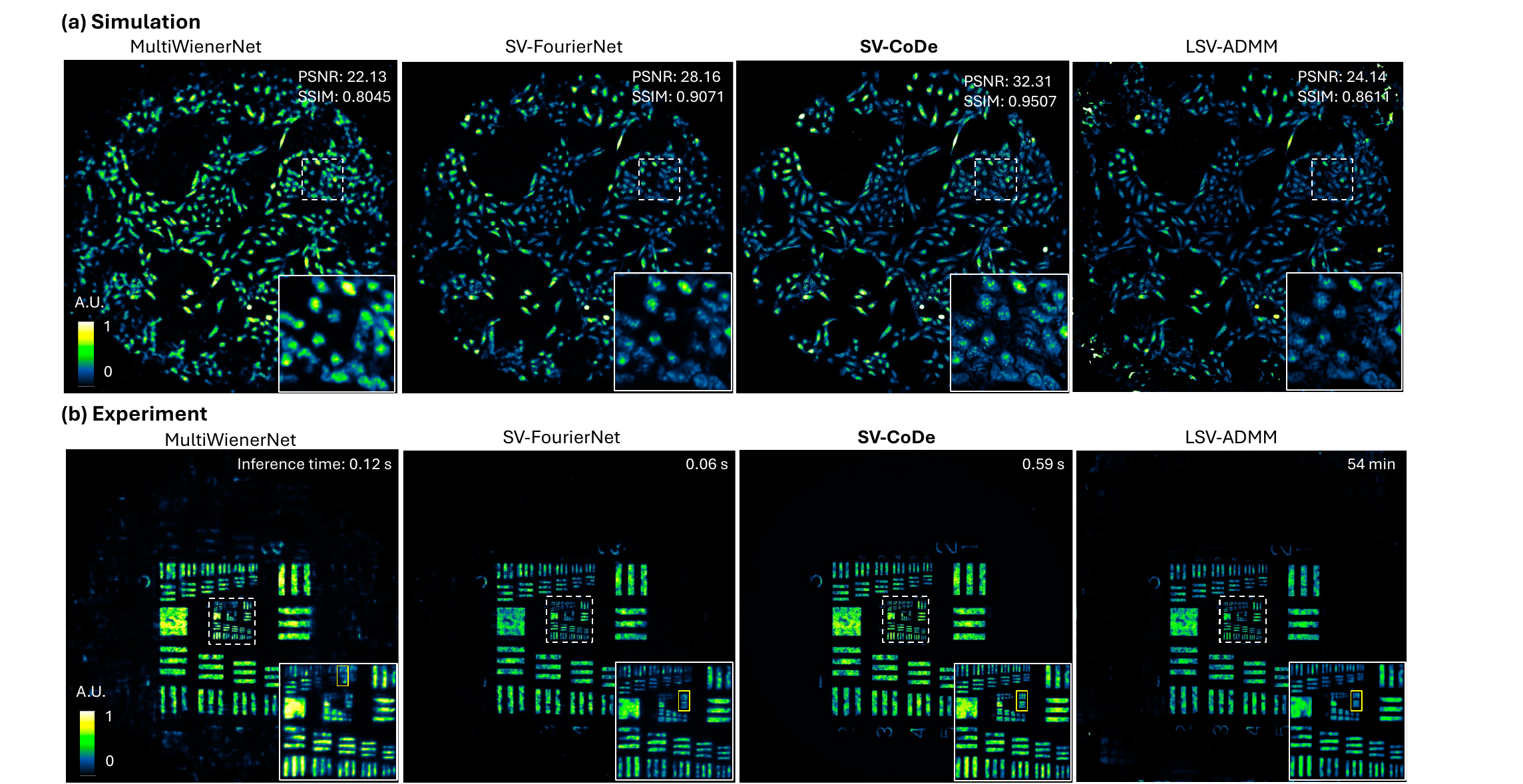}
\caption[Benchmarking results.]{\textbf{Benchmarking results.} (a) \emph{Simulation (cells):} SV-CoDe achieves the highest PSNR/SSIM and produces visibly sharper, cleaner reconstructions than MultiWienerNet and SV-FourierNet. Insets show enlarged regions (white dashed boxes). (b) \emph{Experiment (resolution target):} SV-CoDe, SV-FourierNet, and LSV-ADMM resolve the finest elements (yellow boxes), with SV-CoDe exhibiting higher contrast and fewer artifacts than the other methods. Annotated times report per-frame inference latency on a single GPU.}
\label{fig:benchmark_svcode}
\end{figure}

\noindent To assess reconstruction quality, we benchmark SV-CoDe against MultiWienerNet~\cite{yanny2022deep}, SV-FourierNet~\cite{yang2024wide}, and the model-based LSV-ADMM~\cite{yang2024wide} on both simulated and experimental data (Fig.~\ref{fig:benchmark_svcode}).  

On simulated cells (Fig.~\ref{fig:benchmark_svcode}(a)), SV-CoDe achieves the highest quantitative accuracy, outperforming SV-FourierNet, LSV-ADMM, and MultiWienerNet. Qualitatively, SV-CoDe recovers sharper cellular boundaries and subcellular features with reduced background artifacts, while maintaining fidelity across the full FOV, indicating effective compensation of spatial variance.  

On the experimental resolution target (Fig.~\ref{fig:benchmark_svcode}(b)), SV-CoDe resolves the finest line pairs (yellow boxes) with higher contrast and reduced ghosting artifacts compared to MultiWienerNet, while matching the limiting resolution of SV-FourierNet and LSV-ADMM. Per-frame inference on a single GPU is $0.06$\,s for SV-FourierNet, $0.12$\,s for MultiWienerNet, and $0.59$\,s for SV-CoDe; the iterative LSV-ADMM is orders of magnitude slower.  

Although SV-CoDe has the highest latency among the non-iterative methods, its training and inference operate on independent sub-FOVs, enabling straightforward data parallelism with minimal communication overhead. This design suggests that substantial, near-linear throughput gains are achievable given sufficient multi-GPU compute and I/O bandwidth.  

Overall, SV-CoDe delivers state-of-the-art reconstruction quality with uniform performance across the FOV, while retaining a lightweight, SV parameterization that significantly reduces memory usage relative to prior SV networks.

\subsection{Ablation Study of SV-CoDe Components}
\label{subsec:ablation}
\begin{figure}[t!]
\centering\includegraphics[width=\linewidth]{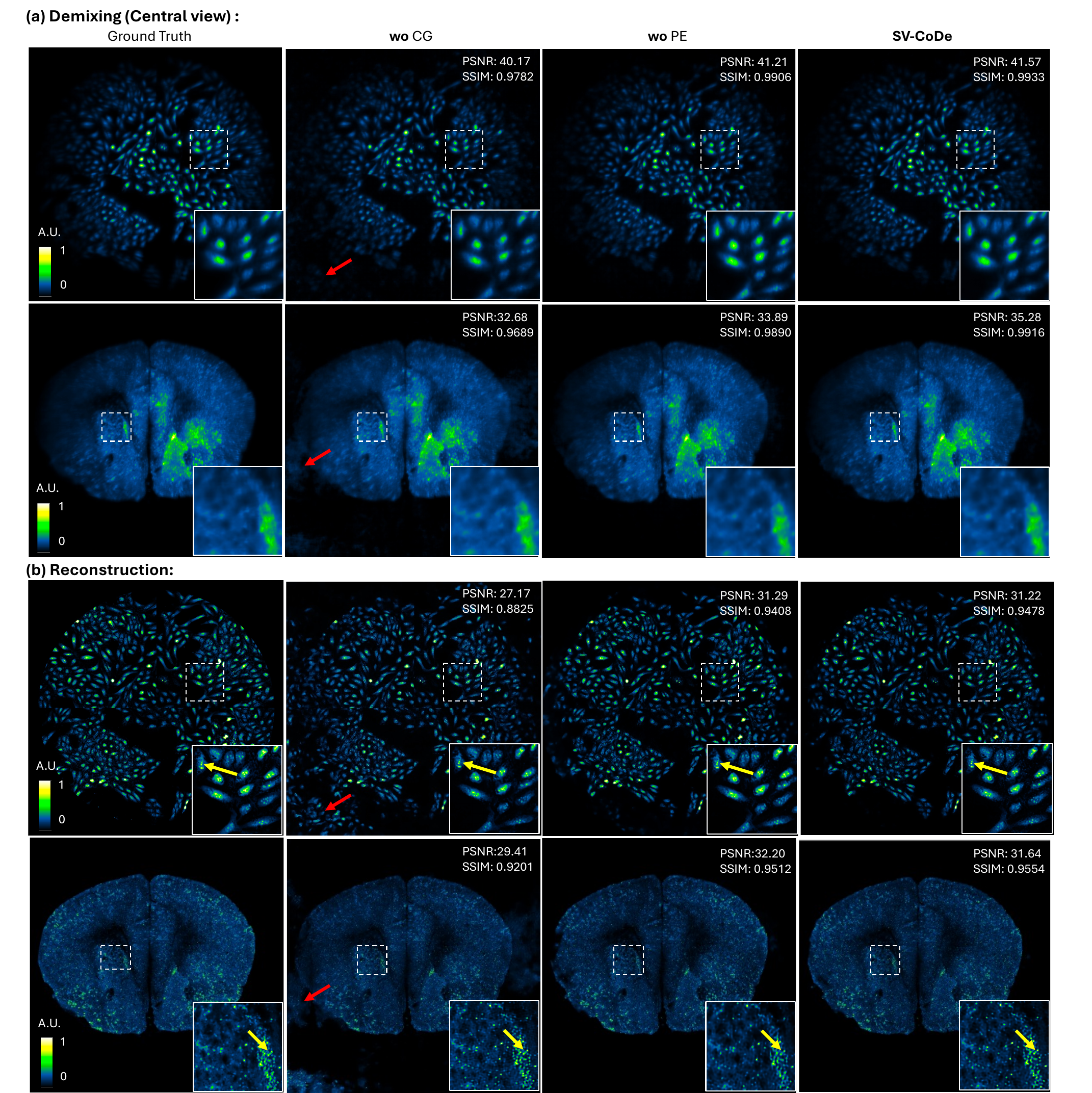}
\caption[Ablation on coordinate gate (CG) and positional encoding (PE).]{\textbf{Ablation on coordinate gate (CG) and positional encoding (PE).} (a) \emph{Central-view demixing.} Columns show the ground truth, w/o CG, w/o PE, and the proposed SV-CoDe. Removing CG leaves residual background, particularly near the field edge where spatial variance is strongest due to sensor truncation (red arrows). Both w/o PE and the full SV-CoDe yield comparable demixing quality. (b) \emph{Final reconstruction.} While central-view demixing alone suffers from severe peripheral aberrations and vignetting, the reconstruction stage recovers fine details and effectively extends the usable FOV. Models w/o CG propagate demixing artifacts into the reconstruction (red arrows), whereas removing PE leads to visibly reduced resolution due to spectral bias (yellow arrows). By leveraging CG for SV deconvolution, SV-CoDe suppresses peripheral ghosting and achieves the highest visual fidelity, closely matching the ground truth.}
\label{fig:ablation_svcode}
\end{figure}

\noindent
We evaluate the contributions of the two key components of SV-CoDe, coordinate gating (CG) and positional encoding (PE), using simulated data from two representative object classes: a densely labeled cellular cluster and a neuron-labeled brain slice with mesoscale structure. We compare three variants: removing CG (\emph{w/o CG}), removing PE (\emph{w/o PE}), and the full model.  

In the demixing stage (Fig.~\ref{fig:ablation_svcode}(a)), removing CG reduces SV-CoDe to a NAFNet with SI convolutions, resulting in residual background, most prominently near the field edge where PSF variation is strongest due to sensor truncation (red arrows). In contrast, \emph{w/o PE} achieves demixing performance comparable to the full model.  

Differences become more pronounced after reconstruction (Fig.~\ref{fig:ablation_svcode}(b)). Without CG, artifacts introduced during demixing propagate and amplify through view fusion, producing peripheral ghosting and degrading image quality toward the FOV edge (red arrows). Without PE, the network exhibits spectral bias that suppresses fine structures, leading to loss of fine features (yellow arrows). The full SV-CoDe, combining CG and PE, yields the closest agreement with the ground truth across both specimen types, suppressing peripheral artifacts while preserving high-frequency detail and uniform image quality across the FOV.  

Taken together, these results indicate complementary roles for the two components: CG provides the spatial variance necessary to correct field-dependent aberrations, whereas PE enables accurate recovery of high-frequency detail across space. Their combination is therefore essential for high-resolution SV reconstruction over a wide FOV.

\subsection{Experimental Demonstration of Consistent Resolution Across a Wide FOV}

\begin{figure}[t!]
\centering\includegraphics[width=\linewidth]{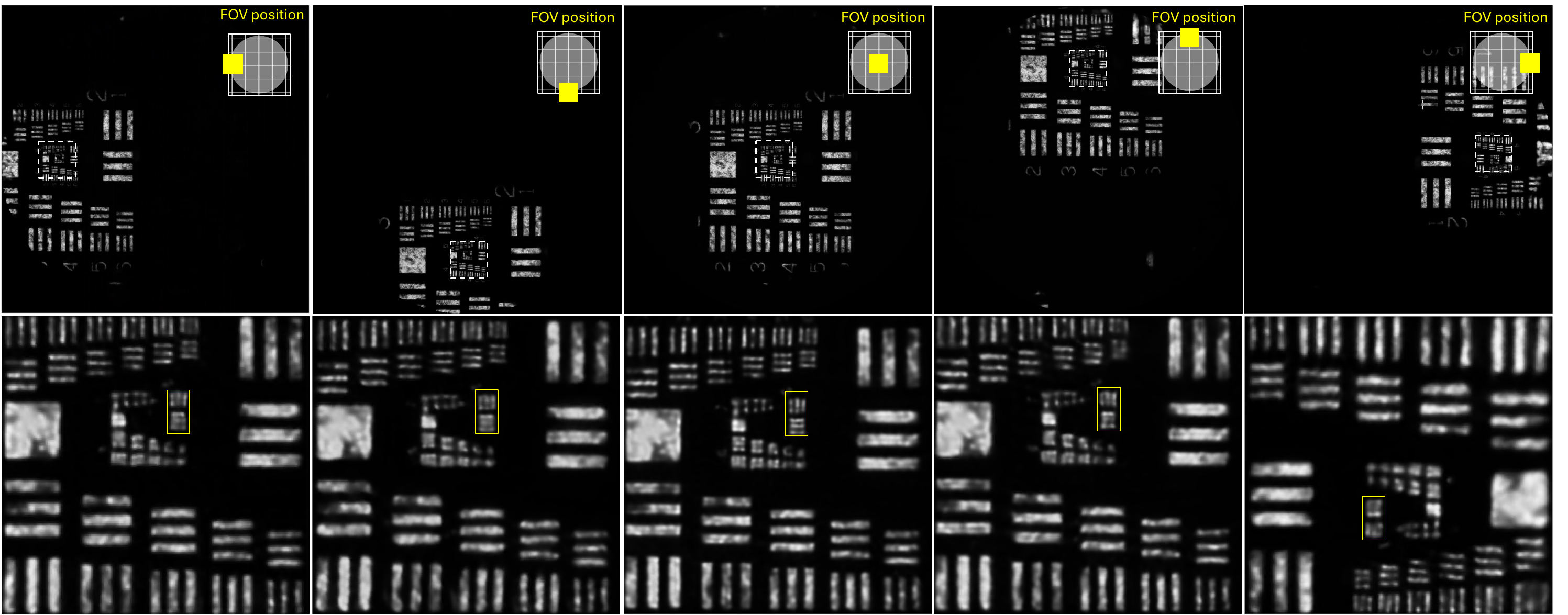}
\caption[Resolution chracterization.]{\textbf{Resolution chracterization.} Reconstructions of a fluorescent resolution target placed at multiple locations within the FOV (insets: yellow square marks the target position). Top row: reconstructions at different field positions. Bottom row: zooms of the white dashed regions; the yellow box marks the finest resolvable line pair. SV-CoDe maintains uniform resolution across different FOV positions.}
\label{fig:uniform_svcode}
\end{figure}

\noindent
To assess resolution uniformity across the FOV, we imaged a fluorescent resolution target at multiple locations spanning both the center and periphery and reconstructed the measurements using SV-CoDe (Fig.~\ref{fig:uniform_svcode}). The top row shows full reconstructions at different field positions, while the bottom row presents zoomed views of the regions indicated by the white dashed boxes. Across all positions, the finest resolvable line pairs (yellow boxes) are identical, corresponding to a consistent $7.8\,\mu\mathrm{m}$ resolution in both vertical and horizontal directions at the center and edge of the FOV. These results demonstrate that SV-CoDe’s SV deconvolution effectively compensates for field-dependent aberrations and vignetting, delivering uniform, high resolution over the entire FOV.

\label{subsec:uniform_fov}

\subsection{Experimental Demonstration on Particle Phantom with Different Densities}

\noindent
We evaluate generalization from simulation to experiment using fluorescent bead phantoms with varying densities. Commercial $10\,\mu$m fluorescent microspheres were dispersed in alcohol to form sparse and dense particle phantoms spanning the full FOV. The raw measurements (insets) exhibit pronounced view multiplexing and field-dependent aberrations.  

\begin{figure}[t!]
\centering\includegraphics[width=\linewidth]{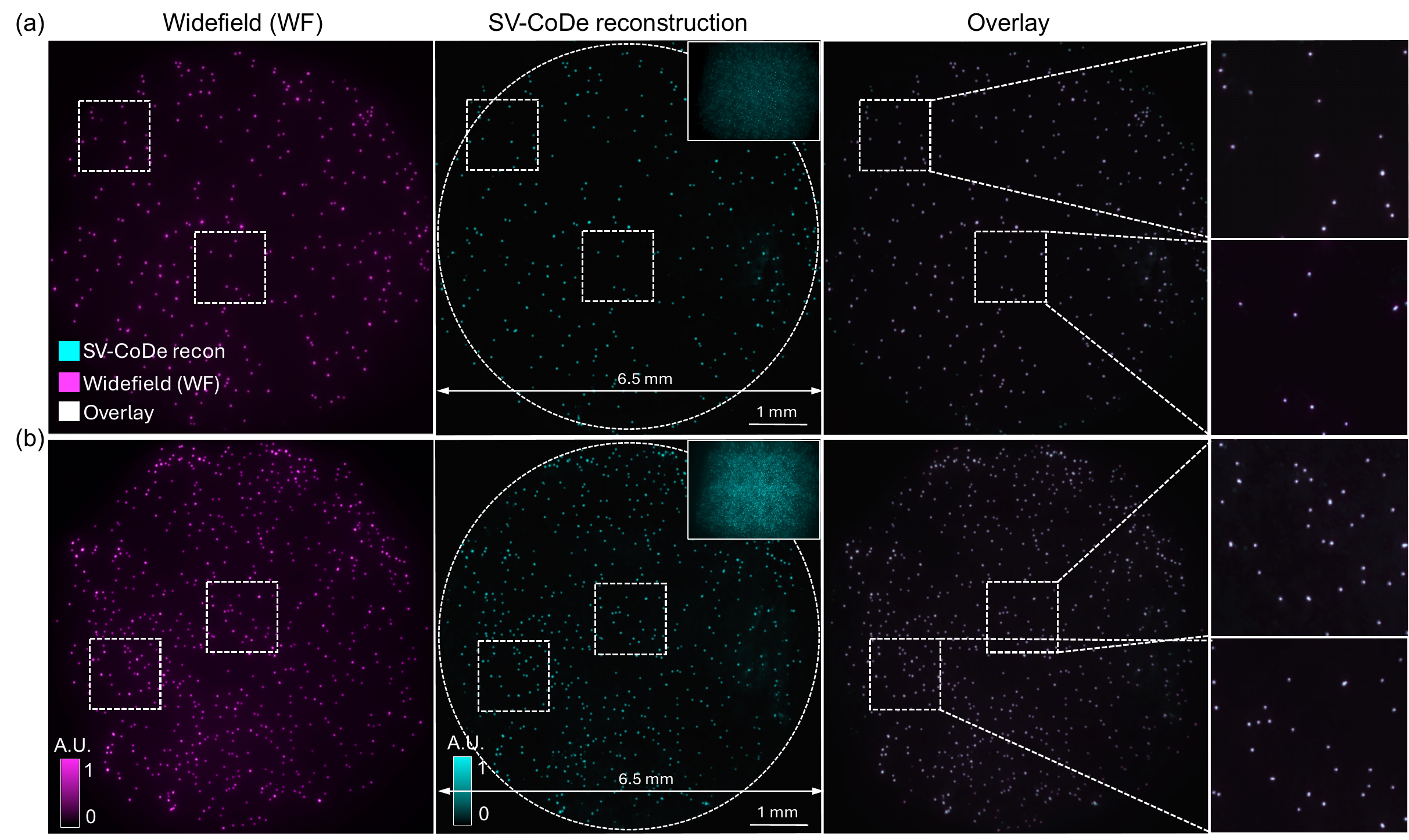}
\caption[Imaging of fluorescent particles with different densities.]{\textbf{Imaging of fluorescent particles with different densities.} Fluorescent $10\,\mu$m bead phantoms imaged at two densities (top: sparse; bottom: dense). From left to right: widefield ground truth (magenta), SV-CoDe reconstruction (cyan) from the multiplexed measurement, and overlay. White dashed boxes indicate regions selected for zooms at the FOV center and edge. Across densities and field positions, SV-CoDe accurately recovers bead locations and sizes, demonstrating robust performance across a wide range of object densities.}
\label{fig:bead}
\end{figure}

Figure~\ref{fig:bead}(a) shows results for a sparse phantom (left to right: wide-field ground truth, SV-CoDe reconstruction, and overlay). Zoomed views from the FOV center and edge confirm that reconstructed particles preserve size and shape uniformly, with no observable peripheral degradation. Figure~\ref{fig:bead}(b) presents a denser phantom; despite increased overlap and crosstalk in the raw measurement, SV-CoDe recovers the majority of particles with high fidelity, and overlays remain well aligned. Zoomed regions again demonstrate comparable performance at the center and periphery.  

Together, these results demonstrate that SV-CoDe, trained solely on simulated data, transfers robustly to experimental measurements across a wide range of object densities, maintaining spatial resolution and avoiding density-dependent artifacts.

\subsection{Experimental Demonstration on Weakly Scattering Brain Tissue}

\begin{figure}[t!]
\centering\includegraphics[width=\linewidth]{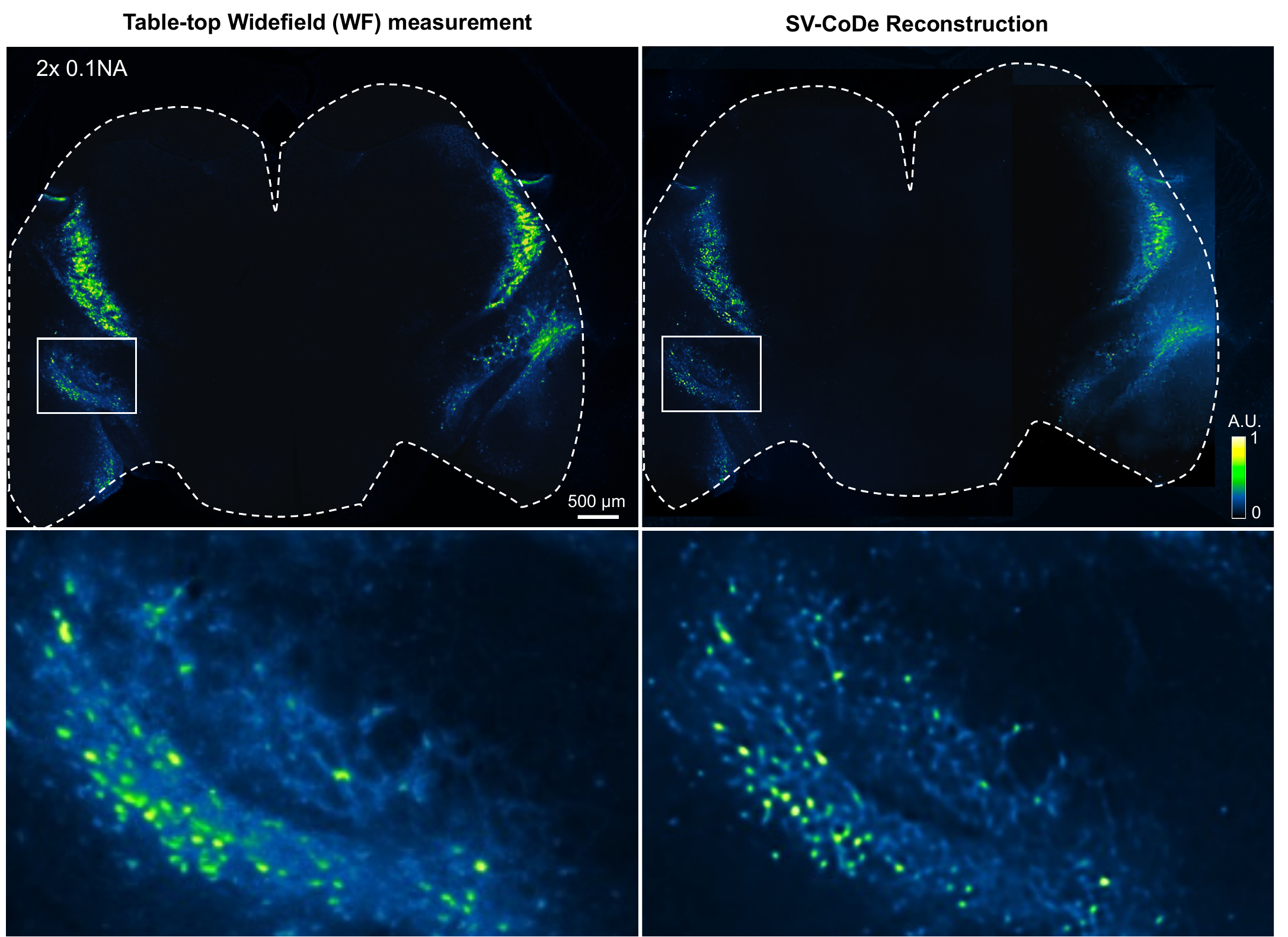}
\caption[Imaging of weakly scattering brain tissue.]{\textbf{Imaging of weakly scattering brain tissue.} Left: table-top widefield (WF, 0.1 NA) image of a fixed mouse brain section with GFP-labeled neurons, serving as ground truth; the dashed line denotes the tissue boundary. Right: SV-CoDe reconstruction from a single multiplexed acquisition, showing close agreement with the WF image over the full FOV. Bottom: zoomed views of the boxed region, where SV-CoDe accurately recovers neuronal structures with enhanced local contrast.}
\label{fig:brain}
\end{figure}

\noindent
We evaluated SV-CoDe on a fixed, weakly scattering mouse brain section with GFP-labeled neurons (Fig.~\ref{fig:brain}). The same section was imaged using a table-top widefield (WF, 0.1 NA) microscope as ground truth. SV-CoDe reconstructs the entire brain section while preserving the bilateral organization of labeled structures and matching the WF resolution. In the zoomed regions, neuronal features align closely with the WF image while exhibiting enhanced local contrast. Despite being trained solely on simulated data, SV-CoDe generalizes robustly to this experimentally challenging, weakly scattering specimen, recovering fine neuronal signals in both dense and sparse regions and maintaining uniform visual quality across the entire tissue.

\subsection{Experimental demonstration on Freely Moving \emph{C.~Elegans}}

\begin{figure}[t!]
\centering\includegraphics[width=\linewidth]{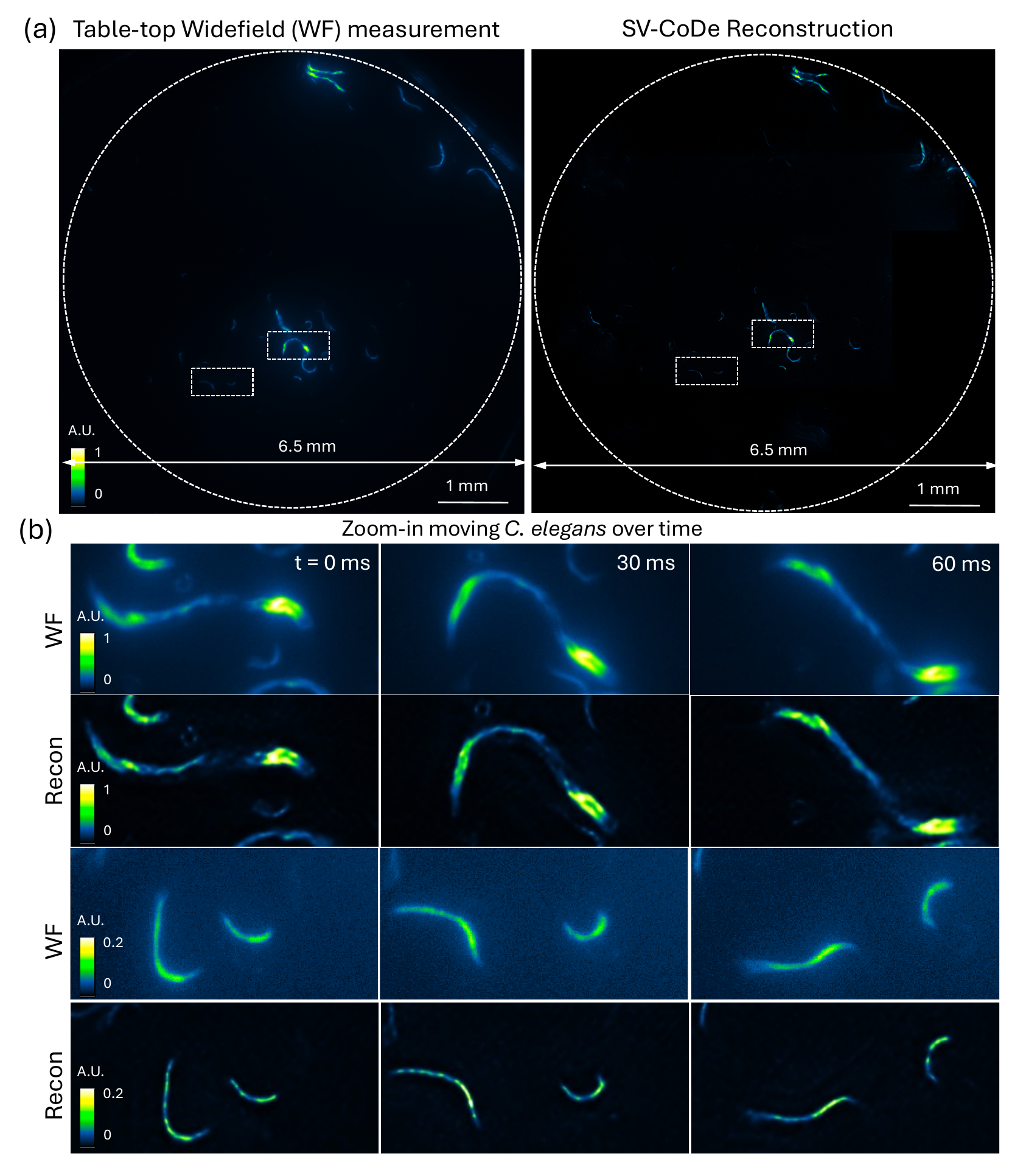}
\caption[Imaging freely moving \textit{C.~elegans}.]{\textbf{Imaging freely moving \textit{C.~elegans}.}
\textbf{(a)} Full-FOV comparison between the table-top wide-field (WF) measurement and the SV-CoDe reconstruction; white boxes indicate regions selected for temporal zooms.
\textbf{(b)} Zoomed sequences at $t{=}0$, $30$, and $60$\,ms for two representative regions. SV-CoDe generalizes robustly to dynamic biological scenes, recovering signals from both high-contrast adult \textit{C.~elegans} and low-contrast juveniles under motion.}
\label{fig:celegans}
\end{figure}

\noindent
We evaluated SV-CoDe on large-scale, dynamic specimens by imaging colonies of freely moving \emph{C.~elegans}. The full-FOV comparison in Fig.~\ref{fig:celegans}(a) shows close agreement between the SV-CoDe reconstruction and the concurrent wide-field (WF) ground truth, with accurate recovery of worms across both central and peripheral regions. Temporal zooms from two representative regions at $t{=}0$, $30$, and $60$\,ms (Fig.~\ref{fig:celegans}(b)) demonstrate robust performance under motion and varying contrast. SV-CoDe resolves worm bodies with enhanced local contrast relative to WF, while reliably capturing dim juvenile animals alongside brighter adults. These results indicate that SV-CoDe, despite being trained solely on simulated data, generalizes robustly to dynamic biological scenes, providing uniform, high-fidelity reconstructions over a wide field.

\section{Conclusion}

\noindent
We presented SV-CoDe, a scalable framework for spatially varying, coordinate-conditioned deconvolution in multi-aperture fluorescence microscopy. The central insight is to replace global, FOV-dependent SV operators with lightweight coordinate-conditioned convolutions, enabling lens-aware adaptation and patch-wise training while decoupling model complexity from FOV size. This design yields uniform, high-resolution reconstruction over a wide FOV with minimal memory and computational overhead.

Extensive ablations and experiments validate the effectiveness of this approach. Coordinate gating provides the spatial variance required to correct field-dependent aberrations and suppress sensor-truncation–induced artifacts, while positional encoding mitigates spectral bias and improves high-frequency recovery. Together, these components form an efficient and physically interpretable operator that captures the dominant SV structure of the imaging system and supports robust reconstruction across the full FOV.

SV-CoDe achieves strong benchmark performance, consistently outperforming prior learning-based SV methods in reconstruction quality while avoiding FOV-dependent parameterizations, and operating orders of magnitude faster than iterative model-based approaches at inference. By decomposing the reconstruction into independent sub-FOVs while preserving global SV behavior, the framework attains high fidelity with improved training efficiency in both data usage and parameter count, making it well suited for high-throughput imaging scenarios.

Despite being trained entirely on physics-based simulations, SV-CoDe generalizes robustly to experimental measurements across diverse samples, including bead phantoms spanning a wide range of emitter densities, weakly scattering brain tissue, and freely moving \emph{C.~elegans}. These results indicate that explicitly embedding SV structure into the network design enables reliable transfer beyond the specific conditions encountered during training.

Looking forward, several extensions are promising. The coordinate-conditioned formulation naturally extends to volumetric and depth-aware reconstruction, offering a path toward scalable 3D SV deconvolution without resorting to global operators. SV-CoDe is also compatible with complementary acquisition strategies, such as event-based sensing~\cite{guo2025dual} or structured illumination~\cite{mertz2011optical}, which may further improve robustness and optical sectioning in low-contrast or highly scattering regimes. More broadly, the CoDe block provides a general abstraction for position-dependent information fusion and may be applicable to other computational imaging systems, such as multi-camera arrays~\cite{brady2012multiscale} and light-field imaging~\cite{ihrke2016principles}. Together, these directions position SV-CoDe as a flexible and physics-aware building block for next-generation wide-field and high-throughput imaging.

\bibliographystyle{IEEEtran}
\bibliography{reference}
\end{document}